\documentclass[twocolumn,11pt]{elsart}
\usepackage{times}
\usepackage[dvips]{graphicx}

\journal{Physics Letters A}

\volume{264}

\firstpage{298}

\pubyear{1999}

\begin{document}

\begin{frontmatter}

\title{Nonlinear Stiffness, Lyapunov Exponents, \\ 
and Attractor Dimension}

\author{Julyan H. E. Cartwright},

\address{Instituto Andaluz de Ciencias de la Tierra, 
CSIC--Universidad de Granada, \\ E-18071 Granada, Spain}

\begin{abstract}
I propose that stiffness may be defined and quantified for nonlinear systems
using Lyapunov exponents, and demonstrate the relationship that exists
between stiffness and the fractal dimension of a strange attractor: that stiff
chaos is thin chaos.
\end{abstract}
\end{frontmatter}

What constitutes a stiff dynamical system? Stiffness is closely related to
numerical methods \cite{cartwright}: the signature of stiffness in a problem is
that, upon integration with a general numerical scheme --- a method not
specially designed for stiff problems --- the routine takes extremely small
integration steps \cite{aiken}, which makes the process computationally
expensive. One view is that stiffness is inextricably linked with the numerical
integration scheme used, so that there would be no such thing as an
intrinsically stiff dynamical system, and the best we could hope for is an
operational definition such as that above \cite{hairer}. Moreover, it has been
proposed that chaotic problems cannot be stiff \cite{corless2}. I argue below
that this is not the case, and provide a definition and a quantitative measure
of stiffness for nonlinear  dynamical systems. I demonstrate how stiffness
affects the geometry of the strange attractor of a chaotic system: that stiff
chaos is thinner --- has smaller fractional part of the fractal dimension --- 
than nonstiff chaos.

When integrating a stiff problem with a variable-step explicit numerical
integration  scheme, the initial step length chosen causes the method to be at
or near  numerical instability, which leads to a large local truncation error
estimate. This causes the numerical routine to reduce the step length
substantially, until the principal local truncation error is brought back
within its prescribed bound. The routine then integrates the problem
successfully, but  uses a far greater number of steps than seems reasonable,
given the smoothness of the solution. Because of this, computation time and
round-off error are a problem when using conventional numerical integration
techniques on stiff problems, and special methods have been
developed for them.

Traditionally in numerical analysis, a linear stiff system of size $n$ is 
defined by \cite{lambert}
\begin{equation}
{\rm Re}(\lambda_i)<0, \qquad 1\leq i\leq n
,\end{equation}
where $\lambda_i$ are the eigenvalues of the Jacobian of the system, with
\begin{equation}
\max\limits_{1\leq i\leq n} |{\rm Re}(\lambda_i)|\gg
\min\limits_{1\leq i\leq n} |{\rm Re}(\lambda_i)|
.\label{lstiff1}\end{equation}
The {\it stiffness ratio}\index{stiffness ratio} $R$ provides a quantitative 
measure of stiffness:
\begin{equation}
R={\frac{\max\limits_{1\leq i\leq n} |{\rm Re}(\lambda_i)|}
{\min\limits_{1\leq i\leq n} |{\rm Re}(\lambda_i)|}}
.\label{lstiff2}\end{equation}
By this definition, a stiff problem has a stable fixed point with eigenvalues 
of greatly differing magnitudes; large negative eigenvalues correspond 
to fast-decaying transients
$e^{\lambda t}$ in the solution. 

As an example of a linear stiff problem, consider the equation
\begin{equation}
y''+1001y'+1000y=0
.\end{equation}
We can write this as the vector equation ${\mathbf y}'=A{\mathbf y}$ where 
\begin{equation}
A=\left(\matrix{0 & 1 \cr -1000 & -1001 \cr}\right)
,\end{equation}
and the eigenvalues are $\lambda_1=-1$ and $\lambda_2=-1000$.
This has solution
\begin{equation}
y=Ae^{-t}+Be^{-1000t}
,\end{equation}
so when integrating the problem numerically with a general variable-step method
we would expect to be able to use a large integration step length after the
$e^{-1000t}$ transient term decays, but in fact the presence of the large
negative eigenvalue $\lambda_2$ prevents this. With  appropriate initial
conditions, one can even set $B=0$ and remove the $e^{-1000t}$ term from the 
solution entirely; one nevertheless has to use a very small step length 
throughout the calculation, as step length is still dictated by the size of 
$\lambda_2$. 

The definition of linear stiffness of Eqs.\ (\ref{lstiff1}) and 
(\ref{lstiff2}) is not relevant for nonlinear systems. The stiffness ratio Eq.\
(\ref{lstiff2}) is often not a good measure of stiffness even for linear
systems, since if the minimum eigenvalue is zero, the problem has infinite
stiffness ratio, but may not really be stiff at all if the other eigenvalues
are of moderate size. The inadequacy of the stiffness ratio is clearly
recognized by numerical analysts, who have moved away from trying to pin down
the definition of stiffness and have generally adopted the pragmatic approach
I alluded to earlier: ``Stiff equations are problems for which explicit methods
don't work'' \cite{hairer}.

\begin{figure}
\includegraphics[width=\columnwidth]{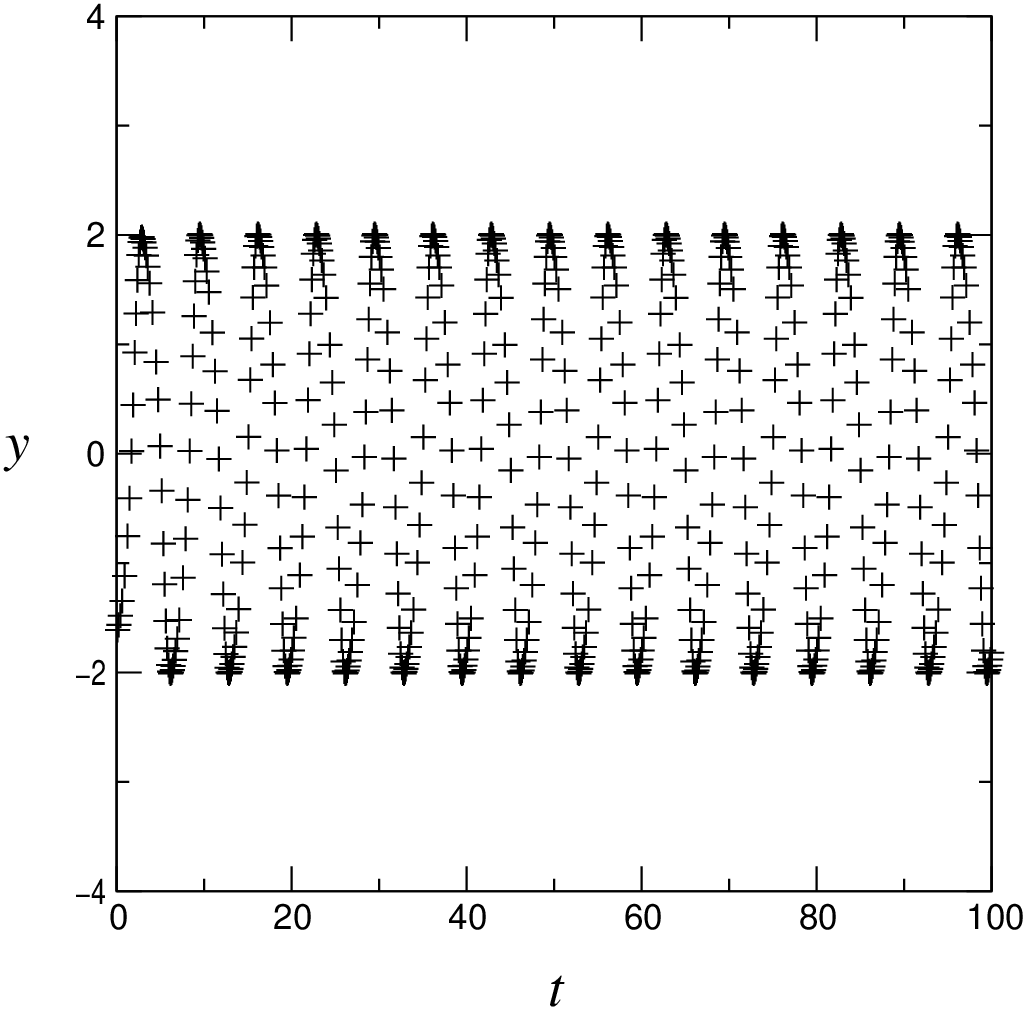}
\includegraphics[width=\columnwidth]{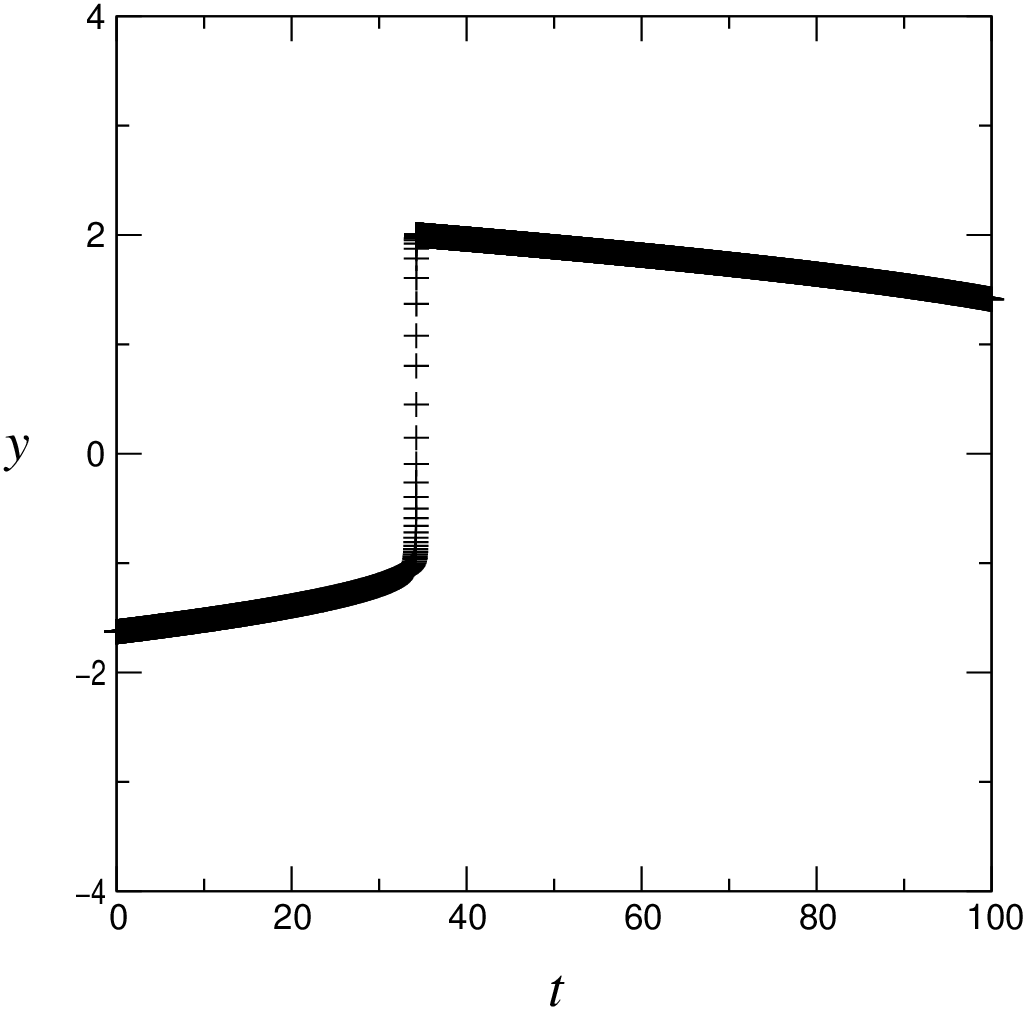}
\includegraphics[width=\columnwidth]{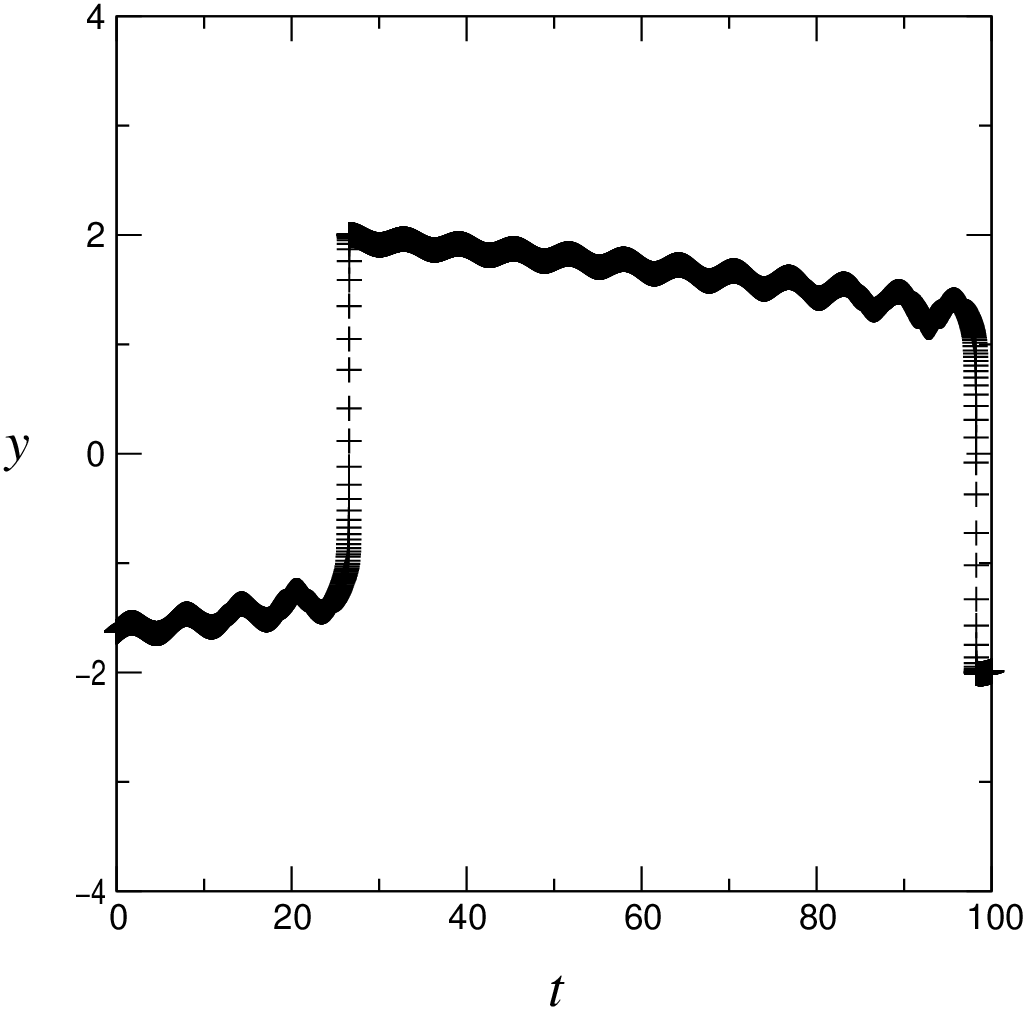}
\caption{\label{vdp}
Numerical integrations using a variable-step fourth-order Runge--Kutta method.
The principal local truncation error is $\varepsilon=10^{-4}$ in all cases.
The van der Pol equation with (a) $\mu=1$ and (b) $\mu=100$;
(c) forced van der Pol equation with $\mu=100$, $A=10$, and $\omega=1$. 
}
\end{figure}

Let us look at a nonlinear stiff problem. 
In Fig.~\ref{vdp}, I show the results of integrating the
van der Pol equation 
\begin{equation}
y''-\mu(1-y^2)y'+y=0
\end{equation}
using a standard variable-step fourth-order Runge--Kutta code \cite{press2}, 
for the cases $\mu=1$
and $\mu=100$. There is the same bound on the principal local
truncation error estimate in both cases. We can see from the far greater
number of steps needed for $\mu=100$, that at large $\mu$ the van der
Pol equation becomes very stiff. The steps are so small at $\mu=100$ that the 
individual crosses representing each step merge to form a continuous broad line 
on the graph. At these large $\mu$ values, the equation
describes a relaxation oscillator \cite{vdp4}. These have
fast and slow states in their cycle which characterizes the jerky motion
displayed in Fig.~\ref{vdp}(b). 
And if we introduce forcing into the van der Pol equation
\begin{equation}
y''-\mu(1-y^2)y'+y=A\cos\omega t
\end{equation}
we obtain a chaotic system \cite{cartlitt,jackson1,tomita,thompson} 
which, like its unforced counterpart, is stiff, as demonstrated in
Fig.~\ref{vdp}(c)), where 
the steps are so small that they have merged together in the image.

Throughout the foregoing, I have illustrated stiff systems with reference to
their effect on the numerical method used for integrating them. Is it possible 
then to define nonlinear stiffness independently of the 
numerical method used? Lambert \cite{lambert} puts forward the definition: 
{\em A system is
said to be stiff in a given interval $x$ if in that interval the neighbouring
solution curves approach the solution curve at a rate which is very large in
comparison with the rate at which the solution varies in that interval.} 
This definition does not make reference to the numerical method used and is
instead concerned with the curvature and local Lyapunov exponents of 
the solution. The curvature of the solution curve 
\begin{equation}
\kappa=\frac{y''}{(1+y'^2)^{3/2}}
\end{equation}
at a point quantifies the wiggliness of the trajectory at that point, while
local Lyapunov exponents are defined as 
\begin{equation}
\gamma_i(\tau,t)=\lim_{\sigma_i(\tau)\to 0}{\frac{1}{\tau}}\ln
{\frac{\sigma_i(t+\tau)}{\sigma_i(t)}}
,\end{equation}
where $\sigma_i(t)$ are the principal axes of an ellipsoidal ball evolving in
time in phase space.
The local Lyapunov exponent $\gamma_i(\tau,t)$ is a generalization of the
Lyapunov exponent $\lambda_i$: the Lyapunov exponent is the limit of $\tau$ 
going to infinity in the local Lyapunov exponent
\begin{equation}
\lambda_i=\lim_{\tau\to\infty}\gamma_i(\tau,t). 
\end{equation}
Lyapunov exponents show the rate of convergence or divergence of neighbouring
trajectories, and an $n$-dimensional system has $n$ Lyapunov exponents 
corresponding to $n$ expanding or contracting directions in phase space.
Since Lyapunov exponents are defined in the infinite time limit,
they cannot reflect differing rates of convergence or divergence in different
parts of a trajectory.
Whereas the Lyapunov exponent is the same for almost all starting points on
a trajectory, the local Lyapunov exponent can vary depending on the starting 
point and the length of trajectory examined.
Fast convergence to a neighbouring trajectory, which implies having large 
negative local Lyapunov exponents, indicates stiffness. 
A system that is equally stiff at all points along a trajectory has a constant 
convergence rate, and in this case, the local Lyapunov exponent will be
the same as the Lyapunov exponent. Often, though, a system can show intervals
of stiff and nonstiff behaviour. Relaxation oscillators like the van der Pol
oscillator are examples of this,
having a stiff slow manifold, and a nonstiff fast manifold. With each 
oscillation we have two intervals of stiff behaviour interspersed with
two intervals of nonstiff behaviour.
 
As it stands, Lambert's definition of stiffness will not do for us, since it 
assumes that the system is not chaotic: the principal local 
Lyapunov exponent is large and negative to obtain fast convergence of 
neighbouring trajectories. Instead, we need to adapt it to allow for chaotic 
behaviour along with stiffness, by looking at the largest negative local 
Lyapunov exponent, rather than the principal local Lyapunov exponent.
Our definition of nonlinear stiffness is then:
{\em A system is stiff in a given interval if in that interval the most 
negative local Lyapunov exponent is large, while the curvature of the 
trajectory is small.} A quantitative measure of nonlinear stiffness at any 
point can then be obtained from the ratio of the most negative local Lyapunov 
exponent and the curvature of the trajectory:
\begin{equation}
R_{nl}=\frac{\left|\min\limits_{1\leq i\leq n}\gamma_i(\tau,t)\right|}
{\kappa(t)}
.\end{equation}
If desired this could be averaged over the trajectory to give a measure of mean
stiffness.

We now have a definition of nonlinear stiffness; what does it imply for chaotic
systems? If a Lyapunov exponent of a system is large and negative, then the 
local Lyapunov exponent must be large and negative at least over some of the 
trajectory, so, given suitable bounds on the curvature of the trajectory, a 
large negative Lyapunov exponent is a sufficient condition for stiffness. On 
the other hand, at least one positive Lyapunov exponent is necessary for chaos. 
Thus for stiff chaos we should have a large spread of Lyapunov 
exponents, with at least one positive, and one large and negative. 

Lyapunov exponents are certainly related to the fractal dimension of an 
attractor. The Kaplan--Yorke conjecture \cite{kaplan} holds that 
the fractal dimension is
\begin{equation}
D_{KY}=\frac{1}{|\lambda_{j+1}|}\sum_{k=1}^j\lambda_k+j,
\end{equation}
where $j$ is the expansion dimension of the system: the largest integer such 
that 
\begin{equation}
\sum_{k=1}^j\lambda_k\geq 0.
\end{equation}
The Kaplan--Yorke estimate is often good --- i.e., very close to the measured
(capacity) dimension. In general the fractal dimension $D$ lies between the 
expansion dimension and the Kaplan--Yorke estimate \cite{grassberger}:
\begin{equation}
j\leq D\leq D_{KY}.
\end{equation}
For example, in a three-dimensional chaotic system, $\lambda_1>0$, 
$\lambda_2=0$, and $\lambda_3<0$, so 
\begin{equation}
D_{KY} = 2+\frac{\lambda_1}{|\lambda_3|}.
\end{equation}
From this, if $|\lambda_3|$ gets larger, with increasing stiffness, 
then the fractal dimension of the strange attractor moves closer to two.

The size of the fractional part of the fractal dimension of a strange 
attractor may be termed the thickness or thinness of the chaos. 
The folded structure of the foliations of the strange attractor --- the 
Abraham and Shaw bagel \cite{abraham} --- is tightly wound in the case of 
thin chaos with small fractional part. On the other hand, if the fractal 
folding of the surface of the strange attractor is macroscopic,
the chaos is thick chaos having a large fractional part to its fractal 
dimension.

\begin{figure}
\includegraphics[width=\columnwidth]{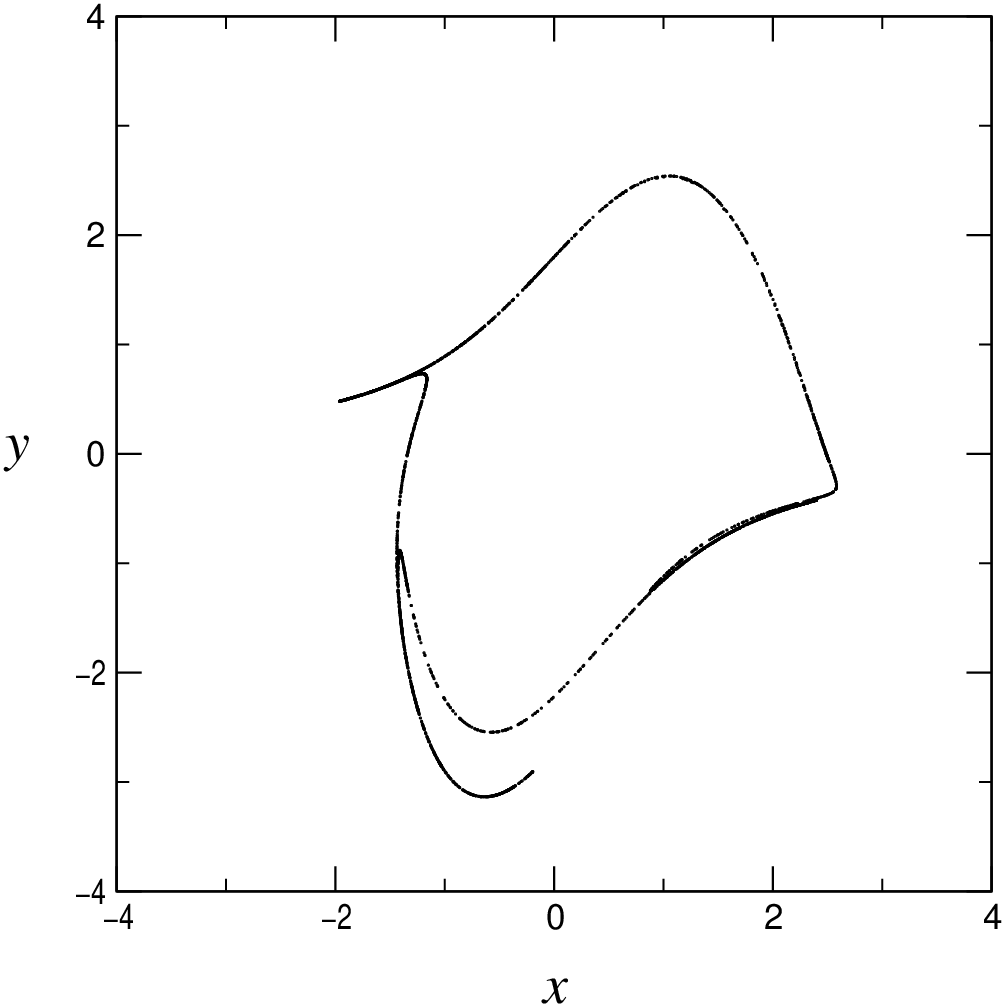}
\includegraphics[width=\columnwidth]{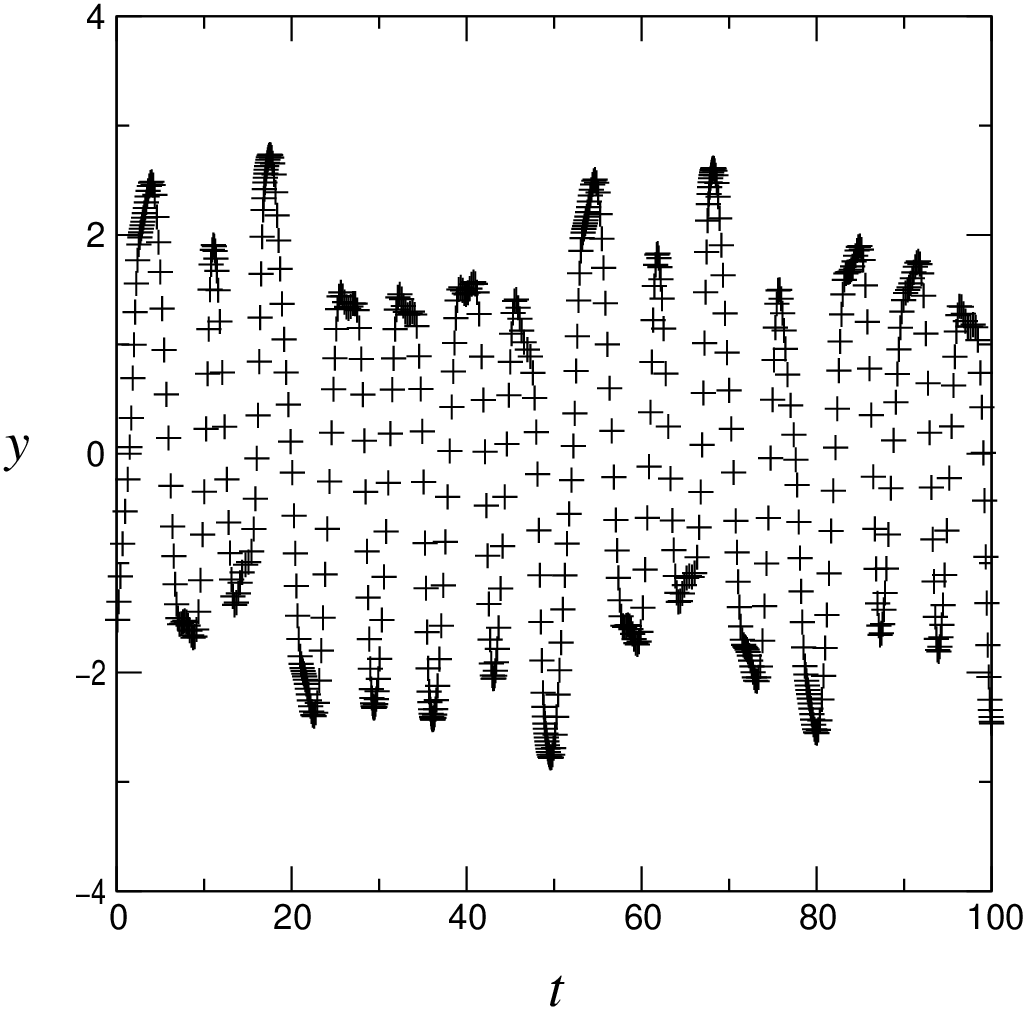}
\caption{\label{shaw}
The Shaw version of the forced van der Pol equation, parameters $A=0.932$,
$\mu=1.18$, and $\omega=1.86$.
(a) Poincar\'e map showing strange attractor.
(b) Numerical integration using a variable-step fourth-order Runge--Kutta
method \cite{press2}, with the same principal local truncation error 
$\varepsilon=10^{-4}$ as in Fig.~\ref{vdp}.}
\end{figure}

We should expect then that stiff chaos will be thin,
and this indeed proves to be the case. 
In Fig.~\ref{vdp} I illustrated the stiffness of the forced van der Pol 
equation. This equation is also well known for the difficulty of capturing 
pictorially its chaotic nature at large $\mu$ \cite{jackson2}.
On the other hand, with the Shaw variant of the forcing \cite{shaw},
which may be written as a three-dimensional autonomous system
\begin{eqnarray}
\dot x&=&-y+\mu(1-y^2)x \nonumber \\ 
\dot y&=&x+A\cos\omega z \nonumber \\
\dot z&=&1
\end{eqnarray}
whose divergence --- the sum of the Lyapunov exponents --- is $\mu(1-y^2)$, 
a strange attractor is easily found at $A=0.932$, $\mu=1.18$, and $\omega=1.86$.
Now, with the benefit of the above analysis, we see how fractal dimension, 
dissipation, and stiffness are all linked together: whereas the normal forced 
van der Pol equation studied at large $\mu$ (high dissipation) is stiff, making
the chaos thin, the attractor in the Shaw variant at small $\mu$ (low 
dissipation) displays thick chaos, and, following these arguments, will not be 
stiff; this is indeed the case (Fig.~\ref{shaw}).

I should like to thank the referees for their helpful comments. I acknowledge
the financial support of the Spanish Consejo Superior de Investigaciones  
Cient\'{\i}ficas.

\bibliographystyle{unsrt}
\bibliography{database}

\end{document}